\documentclass[12pt]{article}

\input epsf.tex
\newcommand{\be}{\begin{equation}}
\newcommand{\ee}{\end{equation}}
\newcommand{\bea}{\begin{eqnarray}}
\newcommand{\eea}{\end{eqnarray}}

\begin{document}

\bigskip 
\begin{titlepage}

\begin{flushright}
UUITP-07/02\\ 
hep-th/0205227
\end{flushright}

\vspace{1cm}

\begin{center}
{\Large\bf Inflation, holography,\\
\smallskip
and the choice of vacuum in de Sitter space\\}

\end{center}
\vspace{3mm}

\begin{center}

{\large Ulf H.\ Danielsson

\vspace{5mm}

Institutionen f\"or teoretisk fysik  \\
Box 803, SE-751 08
Uppsala, Sweden}

\vspace{3mm}

{\tt
ulf@teorfys.uu.se\\}

\end{center}

\vspace{5mm}

\begin{center}
{\large \bf Abstract}
\end{center}
\noindent
A family of de Sitter vacua introduced in hep-th/0203198 as plausible initial conditions for inflation, 
are discussed from the point of view of de Sitter holography. The vacua are argued to be
physically acceptable and the inflationary picture provides a physical interpretation of
a subfamily of de Sitter invariant vacua. Some speculations on the issue of vacuum choice
and the connection between the CMBR and holography are also provided.


\vfill
\begin{flushleft}
May 2002
\end{flushleft}
\end{titlepage}
\newpage 


\section{Introduction}

\bigskip

In recent years there has been a significant interest in the use of
cosmology as a probe of high energy physics. A well established example is
provided by inflation where microscopic quantum fluctuations are magnified
to macroscopic scales, and act as seeds for subsequent structure formation
in the early universe. For a nice review see \cite{liddle}.

An intriguing possibility is that inflation might provide a window towards
physics beyond the Planck scale. In the standard inflationary scenario the
fluctuations start out with a linear size much smaller than the Planck
scale. Nevertheless it is assumed that no new physics appear, and a natural
vacuum for the quantum fluctuations is chosen with this in mind. In several
recent works the rationality of this assumption has been questioned, [2-26].
After all, in the real world we know that fundamentally new physics is to be
expected at the Planck scale and beyond. The idea is that the fluctuation
spectrum might leave an imprint on the CMBR that depends on the details of
this transplanckian physics. Several examples exist in the literature of
specific modifications of the high energy physics which also give a modified
spectrum. In \cite{Danielsson:2002kx} (and more recently in \cite
{Easther:2002xe}) another point of view was taken with the focus on the
choice of vacuum. Initial conditions (for a particular mode) were imposed
when the wavelength was comparable to some fundamental length scale in the
theory. It was shown that this leads to corrections of order $\frac{H}{%
\Lambda }$ to the CMBR-spectrum, where $\Lambda $ is the energy scale of new
physics, e.g. the Planck scale or the string scale, and $H$ is the Hubble
constant during inflation.

The work of \cite{Danielsson:2002kx} only attempted to estimate the natural
size of the corrections but had nothing to say about the actual mechanism
that determines the vacuum. To do that, one needs knowledge about the high
energy physics which goes beyond what is currently available. If one is
looking for large scale effects due to quantum gravity it is tempting to
consider holography. Holography is well established in the AdS setting \cite
{Maldacena:1998re}\cite{Gubser:1998bc}\cite{Witten:1998qj}, while in the
case of de Sitter space there are only more or less well founded guesses of
what to expect. An incomplete list of references is given by [32-57].
Nevertheless, several interesting and suggestive results have been obtained
and the hope is that further work might pin down the principles that governs
the physics of the hologram. But even before this goal is achieved, one can
embark on the project of completing a holographic dictionary where various
aspects of the inflationary theory are expressed in terms of holographic
concepts. In \cite{Strominger:2001gp}\cite{Larsen:2002et}\cite{Halyo:2002zg}
it has been shown how, e.g., inflation with a slowly time varying Hubble
constant (i.e. the slow roll) can be understood as a renormalization group
flow in the hologram.

The main purpose of this note is to translate some of the observations made
in \cite{Danielsson:2002kx} into a language appropriate for the holographic
studies. In particular it is pointed out that the vacua selected in \cite
{Danielsson:2002kx} corresponds to a one-parameter subfamily of the
two-parameter family of de Sitter invariant vacua recently discussed in \cite
{Bousso:2001mw}\cite{Spradlin:2001nb}. The outline of the paper is as
follows. In section two the construction of de Sitter invariant vacua is
reviewed. In section three the connection with \cite{Danielsson:2002kx} is
pointed out. In section four the properties of the vacua relevant for
inflation are discussed. Finally, section five contains some conclusions and
speculations on possible relations with holography.

\bigskip

\section{A de Sitter family of vacua}

\bigskip

De Sitter space does not have a globally time like Killing vector and the
choice of vacuum is therefore a highly non trivial issue \cite
{Gibbons:1977mu}. For the study of physics in de Sitter space there are
several different coordinate systems available. Below we will consider
global coordinates (with spherical spatial sections) that cover the full de
Sitter space, and planar coordinates (with flat spatial sections) that only
cover half of de Sitter space. In the last section we will make use of yet
another type of coordinates, the static coordinates, which are useful when
the focus is on observations made by a particular observer.

The planar coordinates are the ones that are relevant for an inflating
cosmology, where only half of de Sitter space is physical. The Penrose
diagram of de Sitter space has the shape of a square, and can be viewed as
consisting of two parts. An inflating cosmology in the upper right triangle,
and a deflating cosmology in the lower left. They are separated by a light
like surface corresponding to the Big Bang or the Big Crunch respectively.
We will, however, first investigate the choice of vacuum from the point of
view global coordinates, closely following the analysis of \cite
{Bousso:2001mw}\cite{Spradlin:2001nb}.

De Sitter space in $D$ dimensions can be defined through a hypersurface 
\begin{equation}
-X_{0}^{2}+\sum_{i=1}^{D}X_{i}^{2}=\frac{1}{H^{2}},  \label{hyper}
\end{equation}
in $D+1$ dimensional flat Minkowski space. Global coordinates are obtained
by putting 
\begin{eqnarray}
X_{0} &=&H^{-1}\sinh H\tau  \nonumber \\
X_{i} &=&H^{-1}x_{i}\cosh H\tau ,
\end{eqnarray}
where $\sum_{i=1}^{D}x_{i}^{2}=1$. The corresponding metric, in the case of
four dimensional de Sitter space, is given by: 
\begin{equation}
ds^{2}=d\tau ^{2}-H^{-2}\cosh ^{2}H\tau d\Omega _{3}^{2}.
\end{equation}
Following the usual rules of canonically quantizing a massless scalar field,
one finds that 
\begin{equation}
\widehat{\phi }\left( x\right) =\sum_{n}\left[ \phi _{n}\left( x\right) 
\widehat{a}_{n}+\phi _{n}^{\ast }\left( x\right) \widehat{a}_{n}^{\dagger }%
\right] ,  \label{modexp}
\end{equation}
where 
\begin{equation}
\left[ \widehat{a}_{n}^{\dagger },\widehat{a}_{m}\right] =\delta _{nm},
\end{equation}
are creation and annihilation operators, and $\phi _{n}\left( x\right) $ are
modes solving the scalar field equation of motion with $n$ collectively
referring to the discrete quantum numbers of the spherical harmonics on $%
S^{3}$. For clarity we will put an hat on all operator valued fields. The
vacuum of the theory is defined to obey 
\begin{equation}
\widehat{a}_{n}\left| \Omega \right\rangle =0.
\end{equation}
The vacuum in de Sitter space is not unique, instead, as was first discussed
in \cite{chernikov}, and later in \cite{Mottola:ar}\cite{Allen:ux}\cite
{Floreanini:1986tq}, there is a whole family of possible vacua. More recent
discussions, in the context of de Sitter holography, can be found in \cite
{Bousso:2001mw}\cite{Spradlin:2001nb}. Different choices correspond to
different mode expansions, like the one in (\ref{modexp}), where the
different choices are related through Bogolubov transformations. For our
study of the various vacua it will be convenient to use the Wightman
function defined by 
\begin{equation}
G^{+}\left( x,x^{\prime }\right) =\left\langle \Omega \right| \widehat{\phi }%
\left( x\right) \widehat{\phi }\left( x^{\prime }\right) \left| \Omega
\right\rangle =\sum_{n}\phi _{n}\left( x\right) \phi _{n}^{\ast }\left(
x^{\prime }\right) .  \label{wightman}
\end{equation}
The Wightman function is a useful building block from which the other
Green's functions can be obtained. Through the study of the Wightman
function one can learn about the properties of the vacua that one is
interested in. A vacuum that will play a special role is the \textit{%
Bunch-Davies vacuum} (also known as the Euclidean vacuum) which is obtained
by analytical continuation from the Euclidean sphere. It has a Wightman
function given by 
\begin{equation}
G_{E}^{+}\left( x,x^{\prime }\right) =\left\langle E\right| \phi \left(
x\right) \phi \left( x^{\prime }\right) \left| E\right\rangle =\frac{\Gamma
\left( h_{+}\right) \Gamma \left( h_{-}\right) }{16\pi ^{2}}F\left(
h_{+},h_{-};2;\frac{1+P}{2}\right) ,
\end{equation}
where 
\begin{equation}
h_{\pm }=\frac{3}{2}\pm \sqrt{\frac{9}{4}-m^{2}},
\end{equation}
and $P=P\left( x,x^{\prime }\right) $ \ is the de Sitter invariant distance.
Vacua with this simple space time dependence are called de Sitter invariant
and the Bunch-Davies vacuum is therefore an example of such a vacuum.

As discussed in \cite{Mottola:ar}\cite{Allen:ux} one can construct a family
of de Sitter invariant vacua by a Bogolubov transformation, 
\begin{equation}
\phi _{n}\left( x\right) =A\phi _{n,E}\left( x\right) +B\phi _{n,E}^{\ast
}\left( x\right) ,  \label{bogol}
\end{equation}
where 
\begin{equation}
\left| A\right| ^{2}-\left| B\right| ^{2}=1.
\end{equation}
The transformation that we consider treats all modes in the same way
(independent of $n$) and therefore does not spoil the de Sitter invariance.
According to \cite{Mottola:ar}\cite{Allen:ux} it is possible choose the
Euclidean modes such that 
\begin{equation}
\phi _{n,E}^{\ast }\left( x\right) =\phi _{n,E}\left( x_{A}\right) ,
\label{konj}
\end{equation}
where $x_{A}$ is the antipodal point. The antipodal map is, in terms of the
variables $X_{\mu }$ in (\ref{hyper}), obtained by taking $X_{\mu
}\rightarrow -X_{\mu }$. In terms of global coordinates one finds that $%
x=\left( \tau ,\Omega \right) $ is mapped to $x_{A}=\left( -\tau ,\Omega
_{A}\right) $. $\Omega _{A}$ is the point opposite to $\Omega $ on $S^{3}$.
Note that $P\left( x,x^{\prime }\right) =-P\left( x_{A},x^{\prime }\right) $%
. For easy comparison with the case of planar coordinates below, it will be
useful to write (\ref{bogol}) as 
\begin{equation}
\phi \left( x\right) =A\phi _{E}\left( x\right) +B\phi _{E}\left(
x_{A}\right) .  \label{bogolb}
\end{equation}
Using (\ref{wightman}) one finds, finally, a Wightman function given by 
\begin{equation}
G^{+}\left( x,x^{\prime }\right) =\left| A\right| ^{2}G_{E}^{+}\left(
x,x^{\prime }\right) +\left| B\right| ^{2}G_{E}^{+}\left( x^{\prime
},x\right) +AB^{\ast }G_{E}^{+}\left( x,x_{A}^{\prime }\right) +BA^{\ast
}G_{E}^{+}\left( x_{A},x^{\prime }\right) .  \label{wglob}
\end{equation}
After this warm up, we will now turn to the case relevant for inflation,
i.e. the planar coordinates.

The metric in terms of planar coordinates is given by 
\begin{equation}
ds^{2}=dt^{2}-a\left( t\right) ^{2}d\mathbf{x}^{2},
\end{equation}
where $\ a\left( t\right) =e^{Ht}$, is the scale factor. In terms of the
conformal time $\eta =-\frac{1}{aH}$ the metric becomes 
\begin{equation}
ds^{2}=\frac{1}{H^{2}\eta ^{2}}\left( d\eta ^{2}-d\mathbf{x}^{2}\right) .
\label{confmet}
\end{equation}
The relation to the hypersurface of (\ref{hyper}) is given through 
\begin{eqnarray}
\eta &=&-\frac{1}{H^{2}\left( X_{0}+X_{4}\right) } \\
x_{i} &=&\frac{X_{i}}{H\left( X_{0}+X_{4}\right) }\qquad i=1,2,3,
\end{eqnarray}
with $X_{0}+X_{4}>0$. As discussed in \cite{Spradlin:2001nb} the
Bunch-Davies vacuum has a simple description in planar coordinates. It
corresponds to a situation where there are no incoming particles on $\eta
\rightarrow -\infty $. As discussed above, the inflationary universe can be
viewed as the upper right triangle of the full de Sitter space where $\eta
\rightarrow -\infty $ corresponds to the Big Bang, while $\eta \rightarrow 0$
is $\mathcal{I}^{+}$. The lower half of de Sitter space can be covered if we
also consider positive $\eta $. In particular, both signs of $\eta $ are
allowed in the de Sitter invariant distance, which in planar coordinates
becomes

\begin{equation}
P\left( x,x^{\prime }\right) =\frac{\eta ^{2}+\eta ^{\prime 2}-\left( 
\mathbf{x}-\mathbf{x}^{\prime }\right) ^{2}}{2\eta \eta ^{\prime }}.
\end{equation}
It follows from (\ref{planhyp}) that the antipodal map in planar coordinates
takes $x=\left( \eta ,\mathbf{x}\right) $ into $\overline{x}=\left( -\eta ,%
\mathbf{x}\right) $. That is, a map between the two halves of de Sitter
space. Let us now proceed with the quantization, which is formally very
similar to the analysis in global coordinates.

In planar coordinates a massless scalar field can be decomposed as 
\begin{equation}
\widehat{\phi }\left( x\right) =\int d^{3}k\left[ \phi _{\mathbf{k}}\left(
x\right) \widehat{a}_{\mathbf{k}}+\phi _{-\mathbf{k}}^{\dagger }\left(
x\right) \widehat{a}_{\mathbf{k}}^{\dagger }\right] .  \label{phiexp}
\end{equation}
We now assume that the modes can be written in terms of the modes defining
the Bunch-Davies vacuum through 
\begin{equation}
\phi _{\mathbf{k}}\left( x\right) =A\phi _{\mathbf{k,}E}\left( x\right)
+B\phi _{\mathbf{k,}E}\left( \overline{x}\right) ,
\end{equation}
which are the analogue of (\ref{bogolb}) in planar coordinates. For
convenience we write 
\begin{equation}
\phi _{\mathbf{k,}E}\left( x\right) =\phi _{k,E}\left( \eta \right) e^{i%
\mathbf{k}\cdot \mathbf{x}},
\end{equation}
and demand that a relation analogous to (\ref{konj}) is fulfilled: 
\begin{equation}
\phi _{k,E}\left( \eta \right) =\phi _{k,E}^{\ast }\left( -\eta \right) .
\end{equation}
Finally the Wightman function becomes, in complete analogy with the case of
global coordinates,

\begin{equation}
G^{+}\left( x,x^{\prime }\right) =\left| A\right| ^{2}G_{E}^{+}\left(
x,x^{\prime }\right) +\left| B\right| ^{2}G_{E}^{+}\left( x^{\prime
},x\right) +AB^{\ast }G_{E}^{+}\left( x,\overline{x}^{\prime }\right)
+BA^{\ast }G_{E}^{+}\left( \overline{x},x^{\prime }\right) .  \label{planw}
\end{equation}
We will come back to this expression shortly, to see what it has to tell
about the new vacua.

For a massless scalar the above discussion breaks down due to the presence
of a zero mode.\ As pointed out in \cite{Allen:ux}, it is actually
impossible to construct de Sitter invariant vacua for a massless field if
one restricts the attention to just Fock vacua. This does not, however, mean
that de Sitter invariant vacua does not exist. There are two ways to see
this. Either one goes beyond the Fock vacuum and includes a piece
corresponding to a first quantized particle in one dimension, see \cite
{Kirsten}. Or, as proposed in \cite{Tolley:2001gg}, one makes sure that all
correlation functions that one considers only involve derivatives of the
field. The situation is quite similar to the string world sheet theory where
the coordinate field $X^{\mu }$ is not a conformal field. There we can
nevertheless consider insertions of operators of the form $\partial X^{\mu }$%
, or, with a little more thought, operators of the form $e^{ip_{\mu }X^{\mu
}}$.

In the present paper we will be working in momentum space where the
subtleties of the massless case are irrelevant as long as we stay away from
vanishing momentum. Hence the problems of the zero mode do not appear.

\bigskip

\section{Interpreting the vacua}

\bigskip

In this section we will give an interpretation of a one parameter subfamily
of the two parameter family of vacua described in the previous section. To
be explicit, we have, for a massless field in planar coordinates, 
\begin{equation}
\phi _{k,E}\left( \eta \right) =-\eta H\frac{1}{\sqrt{2k}}e^{-ik\eta }\left(
1-\frac{i}{k\eta }\right) ,
\end{equation}
and new modes given by 
\begin{equation}
\phi _{k}\left( \eta \right) =\frac{-\eta H}{\sqrt{2k}}\left( Ae^{-ik\eta
}\left( 1-\frac{i}{k\eta }\right) +Be^{ik\eta }\left( 1+\frac{i}{k\eta }%
\right) \right) .
\end{equation}
We also need the conjugate momentum which is given by 
\begin{equation}
\pi _{k}\left( \eta \right) =\phi _{k}^{\prime }\left( \eta \right) =\frac{%
ik\eta H}{\sqrt{2k}}\left( Ae^{-ik\eta }-Be^{ik\eta }\right) .
\end{equation}
As reviewed in the previous section, there is (up to an overall phase) a two
parameter family of de Sitter invariant vacua. We will now focus on a
particular one parameter subfamily such that: 
\begin{eqnarray}
A &=&e^{i\left( \theta -\beta \right) }\frac{2\beta -i}{2\beta }  \nonumber
\\
B &=&-e^{i\left( \theta +\beta \right) }\frac{i}{2\beta },  \label{enpar}
\end{eqnarray}
where $\beta $ and $\theta $ (the overall phase) are both real. Note that
definitions are such that $\left| A\right| ^{2}-\left| B\right| ^{2}=1$. The
crucial properties of these vacua, parametrized by $\beta $, is that they
imply the existence of a particular moment in time when there is a very
simple relation between $\phi _{k}$ and $\pi _{k}$. One easily verifies that 
\begin{eqnarray}
\phi _{k}\left( \eta _{k}\right) &=&\frac{-\eta _{k}H}{\sqrt{2k}}e^{i\theta }
\nonumber \\
\pi _{k}\left( \eta _{k}\right) &=&-ik\phi _{k}\left( \eta _{k}\right) ,
\label{initialt}
\end{eqnarray}
where 
\begin{equation}
\eta _{k}=-\frac{\beta }{k}.
\end{equation}
Note that the conformal time $\eta _{k}$ depends on $k$. Since $\eta =-\frac{%
1}{aH}$ we find that 
\begin{equation}
p=\frac{k}{a}=\beta H,
\end{equation}
where $p$ is the physical momentum. If we then identify $\beta =\frac{%
\Lambda }{H},$ we find that the physical momentum, corresponding to the mode 
$k$, is equal to $\Lambda $ at conformal time $\eta _{k}$. The vacua defined
by (\ref{enpar}) therefore have the property that a mode $k$ is created in a
state of the form (\ref{initialt}) at the time $\eta _{k}$ when its physical
size is given by the universal scale $\Lambda $.

There are several ways to characterize a state of the form (\ref{initialt}).
As pointed out in \cite{Danielsson:2002kx}, following \cite{Polarski:1995jg}%
, one can note that it corresponds to a state of minimum uncertainty. By
construction it is also what might be called a local Minkowski vacuum
defined through 
\begin{equation}
\widehat{a}_{\mathbf{k}}\left| \Omega \right\rangle =0,
\end{equation}
where $\widehat{a}_{\mathbf{k}}$ is the annihilation operator appearing in (%
\ref{phiexp}).

The analysis in \cite{Danielsson:2002kx} proceeded further by calculating
the perturbation spectrum that would result from these initial conditions.
In the case of $\frac{\Lambda }{H}\gg 1$ it was shown that the fluctuation
spectrum for this new vacuum becomes 
\begin{equation}
P_{\phi }=\left\langle \left| \phi _{k}\left( \eta \right) \right|
^{2}\right\rangle =\left( \frac{H}{2\pi }\right) ^{2}\left( 1-\frac{H}{%
\Lambda }\sin \left( \frac{2\Lambda }{H}\right) \right) ,
\end{equation}
which shows that the natural size of a transplanckian correction is given by 
$\frac{H}{\Lambda }$ in agreement with the claims in \cite{Easther:2001fi} 
\cite{Easther:2001fz}. In a model of inflation where $H$ is slowly changing,
with a spectrum which is not exactly scale invariant, the correction term
will be very sensitive to $k$ through the dependence of $H$ on $k$. That is,
there will be a modulation of $P_{\phi }$. This could be a rather general
phenomena in models where the initial conditions are set at a particular
scale. The modulation that was found is precisely of the same form as in the
numerical work of \cite{Easther:2001fz} which considered a specific example
of slow roll. In further work by the same authors, \cite{Easther:2002xe},
the analytical approach has been extended to the case of a varying Hubble
constant.

In conclusion we have found a very simple interpretation of a subfamily of
de Sitter invariant vacua. It would be interesting to see whether the same
holds true in general.

\bigskip

\section{Some properties of the de Sitter family of vacua}

\bigskip

In the previous section it was shown that the vacuum discussed in \cite
{Danielsson:2002kx}, which is of interest from the point of view of
transplanckian effects, is among the family of de Sitter invariant vacua
discussed in \cite{Bousso:2001mw}\cite{Spradlin:2001nb}. This is a simple
consequence of the way the initial conditions are chosen with all modes
treated equivalently.

The construction of the vacuum, through the use of initial conditions at a
specific scale, provides a physical way of understanding the nature of the
de Sitter invariant vacua. From the opposite point of view, the analysis of
section 2 can give information about the properties of these vacua and tell
whether they make physical sense or not. There are three points to have in
mind in this context.

\textbf{1.} The commutator is independent of the choice of vacuum and the
same is true for the retarded and advanced Greens functions. As a
consequence the classical evolution remains the same. However, if we
consider the Wightman function, or the Feynman propagator, there will be new
singularities on the light cone of the image source at $x_{A}$. It is
therefore tempting to exclude these new vacua as recently discussed in,
e.g., \cite{Tolley:2002cv}. But it is easy to see that the singularities
only appear for points outside of their respective de Sitter horizons. Local
measurements within the horizon will therefore not encounter any peculiar
singularities. The main effect is simply that effectively new initial
conditions on $\eta \rightarrow -\infty $ have been imposed using an image
charge formally positioned on the wrong half of de Sitter.

\textbf{2.} The various de Sitter invariant vacua are physically different.
An example of such a difference is that only the Bunch-Davies vacuum is
thermal. This was discussed in \cite{Bousso:2001mw} where it was shown that
a freely falling detector will only come into thermal equilibrium in the
Bunch-Davies vacuum. It is also quite obvious from the explicit expressions
for the Wightman function that it does not have the analyticity properties
characteristic for a thermal system.

\textbf{3.} At high energies (short distances) the de Sitter invariant vacua
(except the Bunch-Davies) are in general different from ordinary Minkowski
space (even if there are no new singularities). This is expected and
actually the whole point with the reasoning at the end of the previous
section. It simply means that we encounter transplanckian effects when we
probe short distances. The standard argument would have been that the Greens
function should approach the usual Minkowski expression when the distance is
small enough compared to the Hubble scale. However, for small distances we
encounter new physics and there is no regime where the Greens function will
take the Minkowski form.

To conclude, there is no reason of principle to limit one self to the
Bunch-Davies vacuum, at least if one is interested in an inflating cosmology
where only half of de Sitter space is relevant.

\bigskip

\section{Conclusions and speculations on holography}

\bigskip

In this paper we have observed that the vacua discussed in \cite
{Danielsson:2002kx} are among the family of de Sitter invariant vacua
explored in \cite{Bousso:2001mw}\cite{Spradlin:2001nb}. We have also pointed
out that there is no obvious reason to exclude these vacua from a physical
point of view, and a better knowledge of high energy physics is needed in
order to make the correct choice. It would be interesting if the question of
how the vacuum is determined could be made in a holographic setting. In this
respect one should note that the holographic counterpart of the
CMBR-fluctuation spectrum is given by two point correlator in the
holographic theory.

In the by now standard AdS/CFT holographic correspondence, the hologram is
situated at the boundary of AdS. The boundary has a Lorentzian signature and
the theory living there is in many cases a fairly standard CFT. A similar
relation is supposed to hold also in the de Sitter case, where the boundary
is either in the infinite future, $\mathcal{I}^{+}$, or in the infinite
past, $\mathcal{I}^{-}$. A crucial difference from the AdS is that the
boundary has an Euclidean signature. Since we are interested in applying the
construction to inflation we will focus on $\mathcal{I}^{+}$. As we approach 
$\mathcal{I}^{+},$ as $\eta \rightarrow 0,$ it therefore becomes
inappropriate to divide the field according to positive and negative
frequency, see \cite{Das:2002he}. Rather we should distinguish between the
two possible asymptotic behaviors with $\phi \sim \eta ^{h_{\pm }}$,
implying that there are actually two operators in the boundary conjugate to
a given bulk field. We denote these operators by $\mathcal{O}_{\mathbf{k}%
}^{+}$ and $\mathcal{O}_{\mathbf{k}}^{-}$ in momentum space. For the two
point functions one therefore finds that 
\begin{equation}
\left\langle \widehat{\phi }_{\mathbf{k}}\left( \eta \right) \widehat{\phi }%
_{-\mathbf{k}}\left( \eta ^{\prime }\right) \right\rangle =\eta ^{h_{-}}\eta
^{\prime h_{-}}\left\langle \mathcal{O}_{\mathbf{k}}^{-}\mathcal{O}_{-%
\mathbf{k}}^{-}\right\rangle +\eta ^{h_{+}}\eta ^{\prime h_{+}}\left\langle 
\mathcal{O}_{\mathbf{k}}^{+}\mathcal{O}_{-\mathbf{k}}^{+}\right\rangle \sim
\left\langle \mathcal{O}_{\mathbf{k}}^{-}\mathcal{O}_{-\mathbf{k}%
}^{-}\right\rangle ,
\end{equation}
with $h_{+}=3$ and $h_{-}=0$ for a massless bulk field, as we approach the
boundary. It follows, therefore, that the fluctuation spectrum relevant for
the CMBR is, from a holographic point of view, directly given by the
correlator $\left\langle \mathcal{O}_{\mathbf{k}}^{-}\mathcal{O}_{-\mathbf{k}%
}^{-}\right\rangle $. The main observation is that the holographic screen on 
$\mathcal{I}^{+}$ corresponds to the frozen modes that have expanded outside
of the inflationary horizon. If inflation never stopped, they would not be
associated with anything measurable. In \cite{Witten:2001kn} such quantities
on $\mathcal{I}^{+}$ were referred to as \textit{meta-observables}. However,
in the real universe, the modes re-enter through the horizon after the end
of inflation and act like seeds for the fluctuation spectrum of the CMBR.
The study of the CMBR therefore corresponds to a study of the theory on $%
\mathcal{I}^{+}$, and the meta-observables actually turns into real
observables accessible to measurements. This picture, however, breaks down
for small scales on $\mathcal{I}^{+}$, corresponding to late times, when
inflation turns off.

The sensitivity to the initial conditions of inflation explored in \cite
{Danielsson:2002kx} has a precise counterpart in the dependence on the
vacuum of the holographic correlation functions as discussed in \cite
{Bousso:2001mw}\cite{Spradlin:2001nb}, and the question arises what
holography might have to say about the choice of vacuum. The family of de
Sitter invariant vacua all correspond to scale invariant vacua in the theory
on $\mathcal{I}^{+}$. Thinking in terms of planar coordinates, identifying
the comoving momentum $\mathbf{k}$ with the Euclidean momentum in the
hologram, the time evolution (in conformal time) corresponds to an inverted
renormalization group flow from large to small scales. As the flow proceeds,
new modes are integrated in requiring initial conditions. Clearly this
happens, for some particular $\mathbf{k}$, when the corresponding wavelength
becomes of the same order as the cutoff.

To proceed further we need to know a bit more about the de Sitter
holography. A useful coordinate system that brings the focus on what a
particular observer will experience, is the static coordinate system. (A
discussion of the relation between planar and static coordinates from the
point of view of dS/CFT can be found in\cite{Danielsson:2001wt}). The metric
in static coordinates is given by 
\begin{equation}
ds^{2}=-\left( 1-H^{2}r^{2}\right) dt^{2}+\frac{dr^{2}}{1-H^{2}r^{2}}%
+r^{2}d\Omega _{d-2}^{2},  \label{cylmetric}
\end{equation}
where $r<1/H$ corresponds to the \textit{causal diamond} \cite{Bousso}, i.e.
the region of space time which the observer can influence or get influenced
by. If we use planar coordinates such that 
\begin{equation}
ds^{2}=\frac{1}{H^{2}\eta ^{2}}\left( -d\eta ^{2}+d\rho ^{2}+\rho
^{2}d\Omega _{d-2}^{2}\right) ,
\end{equation}
the coordinate transformation that takes us between the planar and static
coordinates is given by 
\begin{eqnarray}
\rho &=&\rho \left( t,r\right) =-Hr\eta  \label{placya} \\
\eta \left( t,r\right) &=&-\frac{e^{-Ht}}{H\sqrt{H^{2}r^{2}-1}}.
\label{placyb}
\end{eqnarray}
In planar coordinates it follows from the metric (\ref{confmet}) that a
source at $\eta $ will give rise to a blob of size $\Delta x=\left| \eta
\right| $ in the boundary at $\mathcal{I}^{+}$. It is therefore natural to
associate a cutoff of order $\left| \eta \right| $ to a spacelike section at
conformal time $\eta $ (the same all over $\mathcal{I}^{+}$). If we now make
a coordinate transformation on the boundary, mapping the plane into a
cylinder, several things will happen. All circles, independent of their
radius $\rho $ in planar coordinates, will be mapped onto the circumference
of a cylinder with a radius independent of the direction $t$ along the
cylinder. If we want a cutoff that is the same all over the cylinder (rather
than all over the plane) we need to change the original cutoff in order to
keep $\rho /\eta $ constant. This will imply a constant $r$ through (\ref
{placya}), with a cutoff on the cylinder that is proportional to $1/r$. From
(\ref{placyb}) we can read off what a particular cutoff corresponds to in
terms of $\eta $. We find that late times (large $t$) along the cylinder
corresponds to small $\eta $.

The cylinder corresponds to a surface of radius $r>1/H$ outside the
cosmological horizon, but if we want the theory to precisely describe the
degrees of freedom on the horizon we need to take $r=1/H$. It follows from,
e.g., (\ref{placya}) that the cutoff now is comparable to the radius of the
cylinder, and there will be only one lattice point encoding all the degrees
of freedom of the whole horizon. Luckily, due to the large central charge
the theory has many degrees of freedom living at each of these lattice
points. This analysis is in line with the analysis in \cite{Kabat:2002hj}.

The above discussion makes it clear that the horizon has, at $\eta $, a size 
$\Delta x=\left| \eta \right| $ implying that the resolution of the hologram
is such that there is only one lattice point per horizon volume. The vacua
that we have been discussing implies that the initial conditions are
imposed, at time $\eta $, on scales given by 
\begin{equation}
\Delta x=\frac{H}{\Lambda }\left| \eta \right| ,
\end{equation}
which is considerably \textit{smaller} than the resolution of the
holographic theory. It follows that the choice of vacuum is encoded in the,
say, $N$ degrees of freedom at the lattice point. If, furthermore, $\Lambda $
is related to the Planck scale (which, for fixed $H$, also determines the
number of degrees of freedom in the holographic theory), it follows that any
transplanckian effects must be associated with $1/N$ effects in the choice
of the vacuum.

Clearly it would be very interesting if one could find a way to select a
natural vacuum from the point of view of the hologram. A specific problem in
this context is to characterize the minimum uncertainty vacua, discussed in
section 3, from a holographic point of view. We hope to return to these and
related questions in the near future.

\section*{Acknowledgments}

I would like to thank Hector Rubinstein and Alexei Abrikosov for comments
and discussions. The author is a Royal Swedish Academy of Sciences Research
Fellow supported by a grant from the Knut and Alice Wallenberg Foundation.
The work was also supported by the Swedish Research Council (VR).

\end{document}